\documentclass{article}

\usepackage{arxiv}

\usepackage[utf8]{inputenc} 
\usepackage[T1]{fontenc}    
\usepackage{hyperref}       
\usepackage{url}            
\usepackage{booktabs}       
\usepackage{amsfonts}       
\usepackage{nicefrac}       
\usepackage{microtype}      
\usepackage{lipsum}
\usepackage{graphicx}
\usepackage{float}
\graphicspath{ {./images/} }

\title{Towards a Universal Measure of Complexity}

\author{
 Jaros{\l}aw Klamut \\
  Faculty of Physics\\
  University of Warsaw,\\
  Pasteur Str. 5\\
  PL-02-093 Warsaw, Poland \\
   \And
 Ryszard Kutner$^*$ \\
  Faculty of Physics\\
  University of Warsaw,\\
  Pasteur Str. 5\\
  PL-02-093 Warsaw, Poland \\
  \texttt{ryszard.kutner@fuw.edu.pl}
  \And
 Zbigniew R. Struzik \\
  Faculty of Physics, University of Warsaw,\\
  Pasteur Str. 5, PL-02-093 Warsaw, Poland \\
  \\
  University of Tokyo, Bunkyo-ku, Tokyo 113-8655, Japan\\
  \\
  Advanced Center for Computing and Communication\\
  RIKEN, 2-1 Hirosawa, Wako 351-0198, Saitama, Japan
}

\begin{document}
\maketitle
\begin{abstract}
Recently it has been argued that entropy can be a direct measure of complexity, where the smaller value of entropy indicates lower system complexity, while its larger value indicates higher system complexity. We dispute this view and propose a universal measure of complexity based on the Gell-Mann's view of complexity. Our universal measure of complexity bases on a non-linear transformation of time-dependent entropy, where the system state with the highest complexity is the most distant from all the states of the system of lesser or no complexity. We have shown that the most complex is optimally mixed states consisting of pure states i.e., of the most regular and most disordered which the space of states of a given system allows. A parsimonious paradigmatic example of the simplest system with a small and a large number of degrees of freedom, is shown to support this methodology. Several important features of this universal measure are pointed out, especially its flexibility (i.e., its openness to extensions), ability to the analysis of a system critical behavior, and ability to study the dynamic complexity.
\end{abstract}


\setcounter{section}{0}
\section{Introduction}\label{section:Introduction}

Analysis of the concept of complexity is a non-trivial task due to its diversity, arbitrariness, uncertainty, and contextual nature \cite{Nicolis,JKSD,Pincus,Dorogov1,AlBer,Dorogov,GrasPro,RichMo,PBEH}. There are many different levels/scales, faces, and types of complexity researched with very different technologies/techniques and tools \cite{XYZ,Wiki1} (and refs. therein). In the context of dynamical systems, Grassberger has suggested \cite{PeGra} that a slow convergence of the entropy to its extensive asymptotic limit is a signature of complexity. This idea was materialized \cite{BNT,PBR} further by information and statistical mechanics techniques.  It generalizes many previous approaches to complexity, unifying physical ideas with ideas from learning and coding theory \cite{Borda}. There is even a connection of this approach to algorithmic or Kolmogorov complexity. The hidden pattern can be an essence of complexity \cite{JPC,GreGol,HGSchus,HHL}. Technics adapted from the theories of information and computation have brought physical science (in particular, the region extended between classical determinism and deterministic chaos) to discovering hidden patterns and quantifying their dynamic structural complexity \cite{Wolfram}.

We must remember that the complexity also depends on the imposed conditions (e.g., boundary or initial) as well as the restrictions adopted. It creates a challenge for every complexity study. It concerns the complexity that can appear in the movement of a single entity and collection of entities braided together. These entities can be as irreducible or straightforward systems as well as complex systems.

When we talk about complexity, we mean irreducible complexity which we can no longer be divided into smaller sub-complexities. We refer to this as a primary complexity. Considering the primary complexity here, we mean one that can be expressed at least in an algorithmic way -- it is an effective complexity if it also contains a logic depth \cite{GMan,GeMan,GeManLl,AMSz,Jaguar}. We should take into account that our models (analytical and numerical) and theories describing reality are not fully deterministic. The evolution of a complex system is potentially multi-branched and the selection of an alternative trajectory (or branch selection) bases on randomly taken decisions.

One of the essential questions concerning a complex system is the problem of its stability/robustness and the question of stationarity of its evolution \cite{RKJM}. What is more, the relationship between complexity and disorder on the one hand, and complexity and pattern on the other are important questions -- especially in the context of irreversible processes, where non-linear ones running away from equilibrium play a central role. Financial markets can be a spectacular example of these processes \cite{Sornet,SorQui,WSGKS,WSzGKS,KDWLGWKS,AJ,AJ2,Ross1,Ross2,Zamb,VSI}.

To the central question: whether entropy is a direct measure of complexity, we answer negatively. In our opinion, the measure of complexity is appropriately, non-linearly transformed entropy. This work is devoted to finding this transformation.

\section{Definition of a universal measure of complexity and its properties}

In this Section, we translate the Gell-Mann general qualitative concept of complexity into the language of mathematics, and we present its consequences.

\subsection{Gell-Mann concept of complexity}\label{section:complexity}

The problem of defining a universal measure of complexity is urgent. For this work, the Gell-Mann concept \cite{GMan,GelMan} of complexity is the inspirational starting point. We apply this concept to irreversible processes, by assuming that both {\em a fully ordered as well as fully disordered systems cannot be the complex}. The fully ordered system essentially has no complexity because of maximal possible symmetry of the system, but the fully disordered system contains no information as it entirely dissipates. Hence, the maximum of complexity should be sought somewhere in between these pure extreme states. This point of view allows introducing a formal quantitative phenomenological complexity measure based on entropy as a parameter of order \cite{Bert,Sornet}. This measure reflects the dynamics of the system through the dependence of entropy on time. The vast majority of works analyzing the general aspects of complexity, including its basis, is based on information theory and computational analysis. Such an approach needs the supplementing with a provision allowing to return from a bit to physical representation -- only then will one allow physical interpretations, including understanding of the causes of complexity.

We define the macroscopic phenomenological partial measure of complexity as a non-linear function of entropy $S$ of the order of $(m,n)$,
\begin{eqnarray}
CX(S;m,n)&\stackrel{\rm def.}{=}&(S^{max} - S)^m(S-S^{min})^n \nonumber \\
&=&CX(S;m-1,n-1)\left[(S-S^{arit})^2-\left(\frac{Z}{2}\right)^2\right],~m,n \ge 1, 
\label{rown:CXS}
\end{eqnarray}
where $S^{min}$ and $S^{max}$ are minimal and maximal values of entropy $S$, respectively, $S^{arit}=\frac{S^{min}+S^{max}}{2}$, and the entropic span $Z\stackrel{\rm def.}{=}S^{max}-S^{min}$, whereas $m$ and $n$ are natural numbers\footnote{An extension to real positive numbers is possible but this is not the case in this paper.}. They define the order $(m,n)$ of the partial measure of complexity $CX$. Let us add that this formula is also met for mesoscopic scale. In other words, complexity appears in all systems for which entropy we can build. Notably, $S^{max}$ does not have to concern the state of thermodynamic equilibrium of the system. It may refer to the state for which entropy reaches its maximum value in the observed time interval. However, in this work, we are limited only to systems having a state of thermodynamic equilibrium. Below we discuss the definition (\ref{rown:CXS}), indicating that it satisfies all properties of the measure of complexity. Of course, when $m=0$ and $n=1$ then $CX$ simply becomes $S-S^{min}$, i.e. the entropy of the system (the constant is not important here). However, when m = 1, n = 0, we get the information contained in the system (constant does not play a role here). Definition (1) gives us a lot more -- showing this is the purpose of this work.

The partial measure of complexity given by Eq. (\ref{rown:CXS}) is determined with the accuracy of the additive constant of $S$ i.e., this constant does not contribute to the measure of the complexity of the system. 

The partial measure of complexity can also be expressed using specific entropy as follows,
\begin{eqnarray}
CX(s;m,n)=\frac{1}{N^{m+n}}CX(Ns;m,n)
\label{rown:specific}
\end{eqnarray}
where $N$ is the number of objects that make up the system and specific entropy $s=S/N$. As one can see, the partial measure of complexity scaling with $N$ is taking place. Thus, the partial measure of complexity we can base both on entropy and specific entropy. The latter approach we can use if we compare the complexities of systems consisting of a different number of objects.

However, often extraction of an additional multiplicative constant (e.g., particle number) to have $s$ independent of $N$, is a technical difficulty or even impossible especially for non-extensive systems. Then, it is more convenient to use the entropy of the system instead of the specific entropy. It is also important to realize that determining extreme entropy values (or extreme specific entropy values) of actual systems can be complicated and requires additional dedicated tools/technologies, algorithms, and models.

The entropy can be here both additive (the Boltzmann-Gibbs thermodynamic one \cite{ETJ} and Shanon information \cite{Borda}) and non-additive entropy (R\'enyi \cite{BeckSchl}, Tsallis \cite{Tsallis}). Apparently, the measure $CX(S)$ is a concave (or convex up) function of entropy $S$, which disappears on the edges at points $S = S^{min}$ and $S = S^{max}$.

It has a maximum
\begin{eqnarray}
CX^{max}=CX(S=S_{CX}^{max};m,n)=m^m~n^n\left(\frac{Z}{m+n}\right)^{m+n}
\label{rown:CXSmax2}
\end{eqnarray}
at point 
\begin{eqnarray}
S = S_{CX}^{max}=\overline{S}=\frac{mS^{min}+nS^{max}}{m+n}=\frac{\frac{1}{m}S^{max}+\frac{1}{n}S^{min}}
{\frac{1}{m}+\frac{1}{n}} 
\label{rown:CXSmax}
\end{eqnarray}
as at this point $\frac{dCX(S)}{dS}\mid _{\overline{S}}~=0$ and $\frac{d^2CX(S)}{dS^2}\mid _{\overline{S}}~<0$. The quantity $CX^{max}$ is well suited for global measurement of complexity because (at a given order $(m,n)$), it depends only on the entropy span $Z$. Perhaps $CX^{max}$ would also be a good candidate for measure of the logic depth of complexity.

\subsection{The most complex structure}

The question now arises about the structure of the system corresponding to entropy $S_{CX}^{max}$ given by Eq. (\ref{rown:CXSmax}). The answer is given by the following constitutive equation,
\begin{eqnarray}
S\left(Y=Y^{CX^{max}}\right)=S_{CX}^{max},
\label{rown:constitS}
\end{eqnarray}
where $Y$ is the set of variables and parameters (e.g., thermodynamic) on which the state of the system depends. However, $Y=Y^{CX^{max}}$ is a set of such values of these variables and parameters which are the solution of Eq. (\ref{rown:constitS}). This solution gives the entropy value $S=S_{CX}^{max}$ that maximizes the measure of complexity, that is $CX=CX^{max}$. Hence, with the value of $Y^{CX^{max}}$ we can finally answer the key question: what structure/pattern is behind it, or how the structure of maximum complexity looks. 

There are a few comments to be made regarding constitutive Equation (\ref{rown:constitS}) itself. It is a (non-linear) transcendental equation in the untangled form relative to the $Y$. This equation should be solved numerically because we do not expect it to have an analytical solution for maximally complex systems. An instructive example of a specific form of this equation and its solution for a specific physical problem is presented in Sec. \ref{section:idgas}.

Eq. (\ref{rown:CXSmax}) legitimizes the measure of complexity we have introduced. Namely, its maximum value falls on the weighted average entropy value, which describes the optimal mixture of completely ordered and completely disordered phases. To the left of $\overline{S}$, we have a phase with dominance of order and to the right a phase with dominance of disorder. The transition between both phases at $\overline {S}$ is continuous. Thus, we can say that the partial measure of complexity we have introduced also defines a certain type of phase diagram in $S$ and $CX$ variables (phase diagram plain). The more detailed information is given in Section \ref{section:entropic}.

\subsection{Evolution of the partial measure of complexity}

Differentiating Eq. (\ref{rown:CXS}) over time $t$ we get the following non-linear dynamics equation, 
\begin{eqnarray}
\frac{dCX(S(t);m,n)}{dt}=\chi _{CX} (S;m,n)\frac{dS(t)}{dt}=(m+n)\left(S_{CX}^{max}-S(t)\right) CX(S(t);m-1,n-1)\frac{dS(t)}{dt},
\label{rown:CXSt}
\end{eqnarray}
where the entropic $S$-dependent (non-linear) susceptibility is defined by
\begin{eqnarray}
\chi _{CX}(S;m,n)\stackrel{\rm def.}{=}\frac{\partial CX(S;m,n)}{\partial S}=(m+n)\left(S_{CX}^{max}-S(t)\right)CX(S(t);m-1,n-1)
\label{rown:CXStd}
\end{eqnarray}
and $\frac{dS(t)}{dt}$ can be expressed, for example, using the right-hand side of the master Markov equation (see ref. \cite{Kamp} for details). However, we must realize that the dependence of entropy on time can be, in general, a non-monotonic, because real systems are not isolated (cf. the schematic plot in Fig. \ref{figure:CXSt}). 
One can see how the dynamics of complexity is controlled in a non-linear way by the evolution of entropy of the system.

In concluding this Section, we state that Eqs. (\ref{rown:CXS}) -- (\ref{rown:CXSt})  provide together technology for studying the multiscale aspects of complexity, including the dynamic complexity. However, it is still a simplified approach as we show in Section \ref{section:conclus}.

\subsection{Properties of the partial measure of complexity}\label{section:arbitZ}

It is worth paying attention to Eqs. (\ref{rown:CXS}) - (\ref{rown:CXSmax}). As one can see, for a fixed span of $Z$ there may be systems of different complexity. In other words, the complexity description only using $CX^{max}$ is insufficient, because there can be many systems with the same entropic span. From this point of view, we assume systems as equivalent, i.e., belonging to the same complexity class $(Z,m,n)$, if they have the same span. However, we can distinguish them as, in general, they differ in the location of $S_{CX}^{max}$. We can say that a given entropic class has greater potential complexity if it has a larger $CX^{max}$. In a given class, the complexity has a larger the system that lies closer to $CX^{max}$, i.e., its current entropy $S$ is closer to $S_{CX}^{max}$. For a given $CX$ with Eq. (1) we get (for each order $(m, n)$) two $S$ solutions: one on the left and the other on the right of $CX^{max}$ (except when $S = S_{CX}^{max}$). That is, all values of the complexity partial measure have doubly degenerated except for $CX^{max}$.

A distinction should be made between two cases of measuring complexity: (i) $Z<m+n$ and (ii) $Z>m+n$.  It is particularly evident when we consider the ratio of both types of complexity measures for $m+n>1$,
\begin{eqnarray}
\frac{CX^{max}(Z_i)}{CX^{max}(Z_{ii})}=\left(\frac{Z_i}{Z_{ii}}\right)^{m+n}<1,
\label{rown:ratiomax}
\end{eqnarray}
where $Z_i$ belongs to case (i) while $Z_{ii}$ to case (ii). As one can see, the greater the exponent $m+n$, the greater the difference between $CX^{max}(Z_{ii})$ and $CX^{max}(Z_i)$.

The alternate form of Eq. (\ref{rown:CXS}),
\begin{eqnarray}
CX(\Delta ) = \left(\frac{n}{n+m} Z +\Delta \right)^n~\left(\frac{m}{n+m} Z - \Delta \right)^m,
\label{rown:CXSDelta}
\end{eqnarray}
where deviation $\Delta =\Delta (t)=S(t)-S_{CX}^{max}$, makes the operating of the $CX$ coefficient easier in the vicinity of $S_{CX}^{max}$, where the parabolic expansion is valid. We have then,
\begin{eqnarray}
CX(\Delta ) \approx CX^{max}\left[1-\frac{1}{2mn}\left((n+m)\frac{\Delta }{Z}\right)^2\right]\approx CX^{max}\exp \left(-\frac{1}{2mn}\left((n+m)\frac{\Delta }{Z}\right)^2\right),  
\label{rown:CXSDelta2}
\end{eqnarray}
that is a Gaussian form, which has variance $\sigma ^2=\frac{nm}{(n+m)^2}Z^2$. As can be seen, only in the narrow range of $S$ around $S_{CX}^{max}$ the measure of complexity $CX$ is symmetrical regardless of the order $(m, n)$.

In fact, only the location of the maximum of $CX(S)$ is determined (for a given range of $S$) by the ratio of $m$ to $n$. However, to have dependence of coefficient $CX$ on entropy in the entire entropy range $S^{min}\leq S\leq S^{max}$, it is necessary to determine two extreme values of entropy ($S^{min}$ and $S^{max}$) and two exponents ($n$ and $m$). In general case, to find these parameters and exponents it is still far from trivial because they have a contextual (and not a universal) character. 

However, in a particular situation, when the maximum complexity is symmetrical, i.e., when $n=m$, we get 
\begin{eqnarray}
S_{CX}^{max}=\overline{S}=\frac{S^{min}+S^{max}}{2} 
\label{rown:SCX2}
\end{eqnarray}
and
\begin{eqnarray}
CX^{max}=\left(\frac{Z}{2}\right)^{2n}. 
\label{rown:CX2}
\end{eqnarray}

 Definition (\ref{rown:CXS}) of the partial measure of complexity applies both to single- and multi-particle issues because entropy can also be built even for a very long single-particle trajectory. Moreover, Definition (\ref{rown:CXS}) emphasizes our point of view that any evolving system for which one can introduce the concept of entropy and which have a state of thermodynamic equilibrium (for which entropy reaches a global maximum), contains at least a signature of complexity. For systems of negligible complexity, i.e., for which $S\approx S^{min}$ or $S\approx S^{max}$, the coefficient $CX(S)$ is close to zero. It does not mean, however, that we cannot locate $S_{CX}^{max}$ near $S^{min}$ or $S^{max}$. It is sufficient then to have strongly asymmetric situations when $n \ll m$ or $n \gg m$, respectively.

\subsubsection{Significant partial measure of complexity}

\begin{figure}[H]
\centering
\includegraphics[width=10cm]{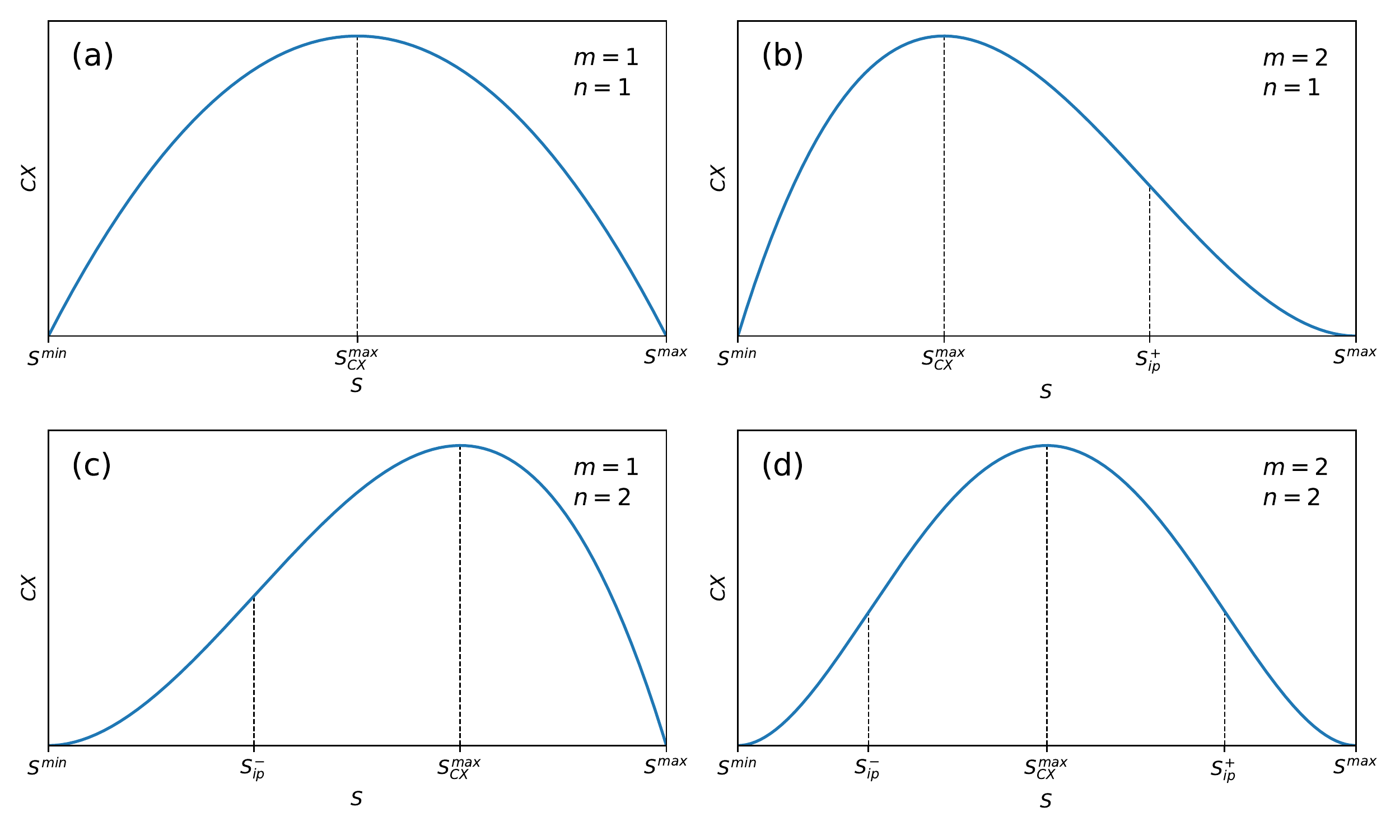}
\caption{Schematic plots of measure of complexity $CX(S;m,n)$ vs. $S$ for four characteristic cases: (\textbf{a}) Case $n=m=1$ where no inflection points, $S_{ip}^{\mp}$ are present. (\textbf{b}) Case  $m=2$ and $n=1$ where a single inflection point $S_{ip}^+$ is present. (\textbf{c}) Case $m=1$ and $n=2$ where a single inflection point $S_{ip}^-$ is present. (\textbf{d}) Case $m=2$ and $n=2$ where both inflection points are present. The shape of the curve, containing two inflection points, is typical for partial measures of complexity, characterized by exponents $m, n\geq 2$.}
\label{figure:CXS}
\end{figure}

We consider the partial measure of complexity to be significant when the entropy of the system locates between two inflection points of the $CX(S;m,n)$ curve, i.e., in the range $S_{ip}^-\leq S \leq S_{ip}^+$. This case occurs for $n,m\geq 2$. We then obtain,
\begin{eqnarray}
S^{min}<S_{ip}^{\mp }= S^{min}+\frac{n(n-1)}{\sqrt{n(n+m-1)}}\frac{S^{max}-S^{min}}{\sqrt{n(n+m-1)}\pm \sqrt{m }}<S^{max}, 
\label{rown;S-+ip}
\end{eqnarray}
see Fig. \ref{figure:CXS}(d) for details. 

There are two different cases where a single inflection point is present. Namely,
\begin{eqnarray}
S^{min}<S_{ip}^- =\frac{2S^{max}+m(m-1)S^{min}}{2+m(m-1)}<\overline{S}, & \mbox{for $m\geq 2$,~$n=1$},
\label{rown:S-ip}
\end{eqnarray}
and
\begin{eqnarray}
\overline{S}<S_{ip}^+ =\frac{2S^{min}+n(n-1)S^{max}}{2+n(n-1)}<S^{max}, & \mbox{for $m=1,~n\geq 2,$}.
\label{rown:S+ip}
\end{eqnarray}
The case defined by Eq. (\ref{rown:S-ip}) we present in Fig. \ref{figure:CXS}(b), while defined by Eq. (\ref{rown:S+ip}) we present in Fig. \ref{figure:CXS}(c).  

For $n=m=1$ the curve $CX(S;m,n)$ vs. $S$ has no inflection points and it looks like a horseshoe (cf. Fig. \ref{figure:CXS}(a)).

Notably, we can equivalently write, 
\begin{eqnarray}
S^{min}<S_{ip}^{\mp }= S^{max}-\frac{m(m-1)}{\sqrt{m(n+m-1)}}\frac{S^{max}-S^{min}}{\sqrt{m(n+m-1)}\mp \sqrt{n}}<S^{max}, & \mbox{for $n,m\geq 2$}. 
\label{rown:AltS}
\end{eqnarray}

Let's consider the span $Z_{ip}=S_{ip}^+-S_{ip}^-$ of the two-phase area. From Eq. (\ref{rown;S-+ip}) or equivalently from Eq. (\ref{rown:AltS}) we obtain that,
\begin{eqnarray}
Z_{ip}= \frac{2\sqrt{nm}}{(n+m)\sqrt{n+m-1}}Z.
\label{rown:Spanip}
\end{eqnarray}
As one can see, span $Z_{ip}$ depends linearly on span $Z$ and in a non-trivial way on exponents $n$ and $m$. Thus, with the $Z$ set, only $Z_{ip}$'s non-trivial dependence on the order $(m,n)$ of measure of complexity $CX$ occurs, which is different from $CX^{max}$ dependence. In other words, $Z_{ip}$ is less sensitive to complexity than $CX^{max}$.

The significant partial measure of complexity ranges between two inflection points only for case $n,m \geq 2$ (cf. Fig. \ref{figure:CXS}(d)). Indeed, a mixture of phases is observed in this area. For areas where $S^{min}\leq S <S_{ip}^-$ and $S_{ip}^+ < S \leq S^{max}$ we have (practically speaking) only single phases, ordered and disordered, respectively (see Section \ref{section:entropic} for details). The case defined  by Eq. (\ref{rown;S-+ip}) and equivalently by Eq. (\ref{rown:AltS}) are the most general that takes into account the fullness of complexity behavior as a function of entropy. Other cases impoverish the description of complexity. Therefore, we will continue to consider the situation when $n,m\geq 2 $.

The choice of any of the $CX(S;m,n)$ forms (i.e. exponents $n$ and $m$) is somewhat arbitrary function of state of the system as it depends on the function of state, that is on the entropy. In our opinion, the shape of the $CX(S;m,n)$ coefficient vs. $S$ we present in Fig. \ref{figure:CXS}(b) is the most appropriate because only then the significant complexity (ranging between $S_{ip}^-$ and $S_{ip}^+$) is well defined. 

In generic case we should use, however, the definition (\ref{rown:CXS}). Then, we can define the order of the partial complexity using the pair of exponents $(n,m)$ . The introduction of the order of the partial complexity is in line with our perception of the existence of multiple levels of (full) complexity.

\begin{figure}[H]
\centering
\includegraphics[width=10cm]{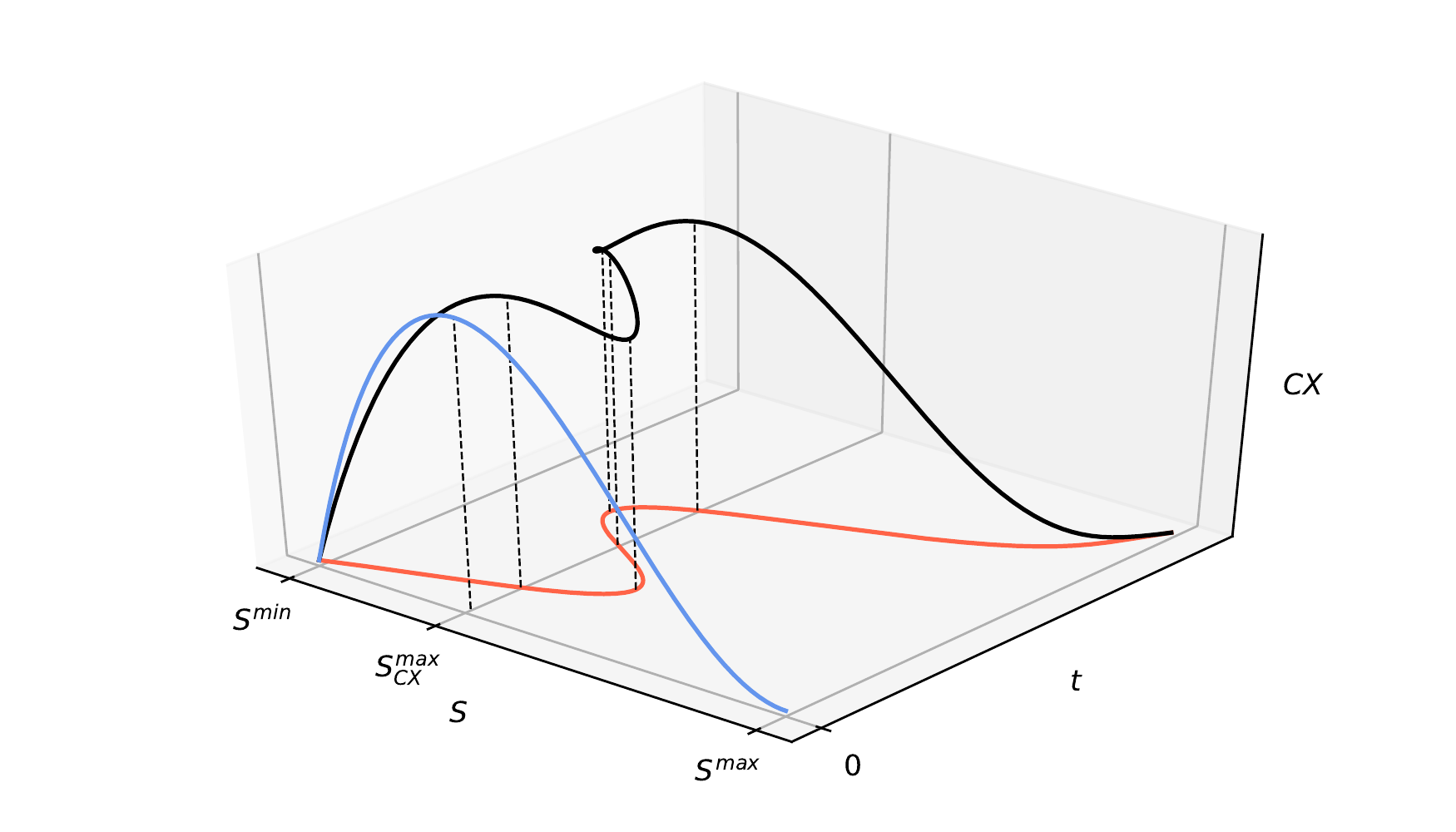}
\caption{Schematic plot of the partial measure of complexity $CX(S;m,n)$ vs. $S$ and $t$. The red curve shows the dependence of entropy $S$ on time $t$. The black curve represents $CX(S(t);m,n)$ in three dimensions. The blue curve represents projection of the black curve on the $(S,CX)$ plane. Indeed, different variants of this blue curve we show in Fig. \ref{figure:CXS}. The non-monotonic dependence of the entropy on time visible here indicates the open nature of the system. However, this non-monotonicity is not visible through the blue curve. For instance, the three local maxima of the black curve, colapses to one of the blue curve.}
\label{figure:CXSt}
\end{figure}

We only discover the nature of the $CX$ factor, i.e. its dynamics and in particular its dynamical structures, when we analyze the entropy dynamics $S(t)$ (see Fig. \ref{figure:CXSt} for details).

The measurability of the partial measure of complexity is necessary to characterize it quantitatively and to be able to compare different complexities. Following Gell-Mann \cite{GelMan}, we must identify the scales at which we perform the analysis and thus determine coarse-graining to define the entropy. Its dependence on complexity cannot be ruled out.

However, the question of direct measurement of the partial measure of complexity in an experiment (real or numerical) remains open.

\subsection{Remarks on the entropic susceptibility}\label{section:entropic}

An essential tool for studying phase transitions is the system susceptibility -- in our case, the entropic susceptibility of the partial measure of complexity. It plays the role of the order parameter here. 

\begin{figure}[H]
\centering
\includegraphics[width=10cm]{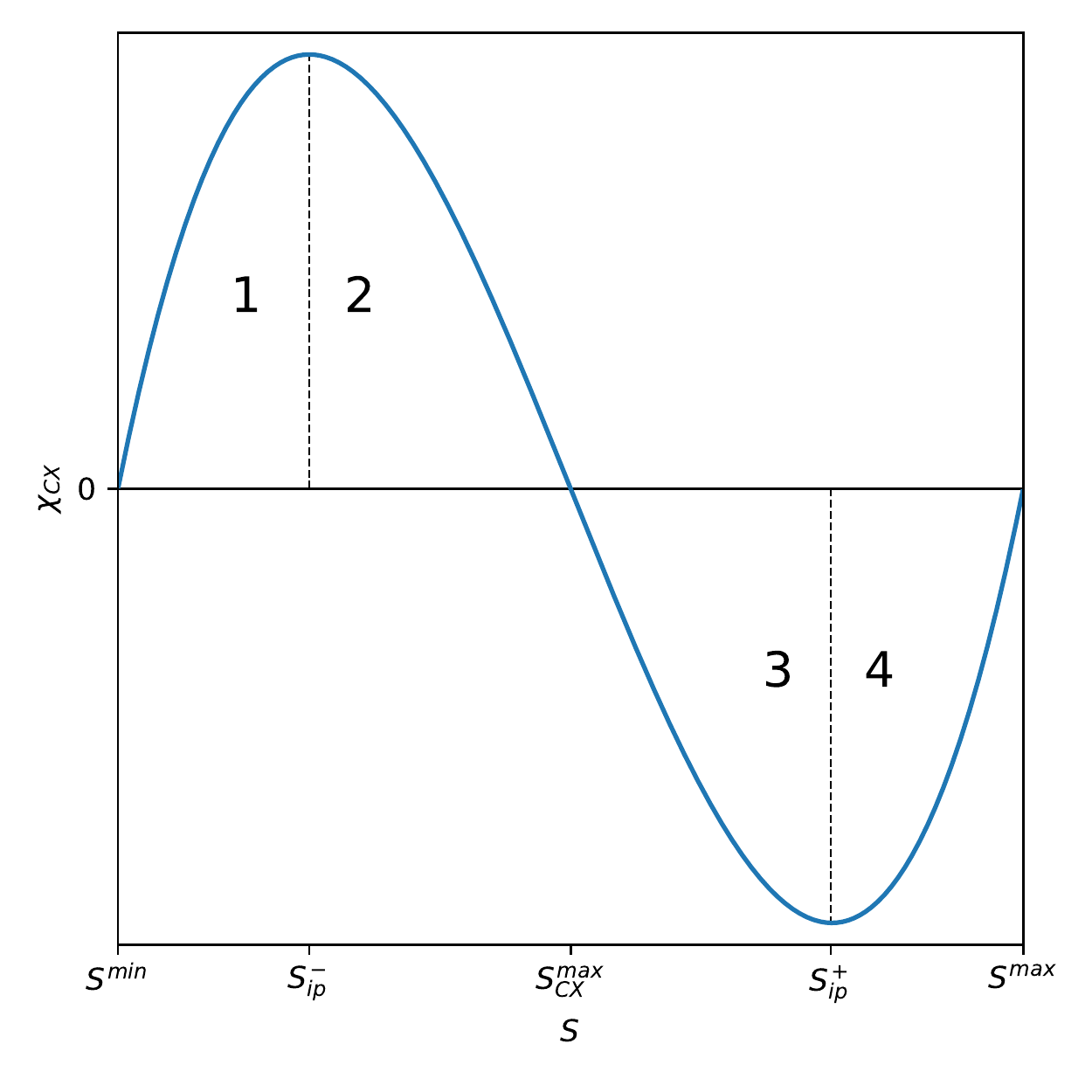}
\caption{Schematic plot of the entropic (non-equilibrium) susceptibility $\chi _{CX}(S;m,n)$ of the partial measure of complexity vs. $S$ at arbitrary fixed order $(m=2,n=2)$. The finite susceptibility value at the $S_-^{ip}$ and $S_+$ phase transition points may be considered to correspond to finite susceptibility value in absorbing non-equilibrium phase transition in the model of direct percolation at a critical point in the presence of an external field \cite{HHL}. However, the situation presented here is richer, because susceptibility changes its sign, smoothly passing through zero at $S=S_{CX}^{max}$. At this point, the system is extremely robust.}
\label{figure:Suscepti}
\end{figure}

The plot of susceptibility $\chi _{CX}$ vs. $S$ is presented in Figure \ref{figure:Suscepti}. Four phases are visible (numbered from 1 to 4).
Phase number 1 is almost entirely ordered -- the disordered phase input is residual. At point $S_-^{ip}$, there is a phase transition to the mixed-phase marked with number 2 still with the predominance of the ordered phase. At the $S_-^{ip}$ transformation point, the entropic susceptibility reaches a local maximum. By further increasing the entropy of the system, it enters phase 3 as a result of phase transformation at the very specific $S_{CX}^{max}$ transformation point. At this point, the entropic susceptibility of the partial measure of complexity disappears. This mixed phase (number 3) is already characterized by the advantage of the disordered phase over the ordered one. Finally, the last transformation which occurs at $S_+^{ip}$, leads the system to the dominating phase by the disordered phase -- the input of the ordered phase is residual here. At this transformation point, the susceptibility reaches a local minimum. Intriguingly, entropic susceptibility can have both positive and negative value passing smoothly through zero at $S=S_{CX}^{max}$ where system is exceptionally robust. The presence of phases with positive and negative entropic susceptibility is an extremely intriguing phenomenon.

It should be emphasized that the values of local extremes of the entropic susceptibility of the partial measure of complexity are finite and not divergent, as in the case of  (equilibrium and non-equilibrium) phase transitions in the absence of an external field. From this point of view, Definition (1) should be treated as a simplified definition of complexity measure. Extending this definition to describe the critical behavior of a system is possible -- however, this is a topic for future work. 

\subsection{Universal full measure of complexity}

Universal the full measure of complexity $X$ is a weighted sum of the partial measures of complexity $CX(S;m,n)$ for individual scales. That is, 
\begin{eqnarray}
X(S;m_0,n_0)=\sum_{m\geq m_0,n\geq n_0}w(m,n)CX(S;m,n),~m_0,n_0\geq 0,
\label{rown:umsX}
\end{eqnarray}
where $w(m,n)$ is a normalized weight, which must be given in an explicit form. This form is to some extent imposed by the power-law form of partial complexity. Namely, we can assume
\begin{eqnarray}
w(m,n)=\left(1-\frac{1}{M}\right)^2\frac{1}{M^{m-m_0+n-n_0}},~M>1,
\label{rown:mn00}
\end{eqnarray}
which seems to be a particularly simple because
\begin{eqnarray}
\frac{w(m+1,n)}{w(m,n)}=\frac{w(m,n+1)}{w(m,n)}=\frac{1}{M},
\label{rown:WNmn}
\end{eqnarray}
independently of $m$ and $n$.

From Eqs. (\ref{rown:umsX}) and (\ref{rown:mn00}) we directly obtain
\begin{eqnarray}
X(S;m_0,n_0)=\left(1-\frac{1}{M}\right)^2\frac{(S^{max}-S)^{m_0}}{1-\frac{S^{max}-S}{M}}\frac{(S-S^{min})^{n_0}}{1-\frac{S-S^{min}}{M}}.
\label{rown:CXmnwN}
\end{eqnarray}
The $M$ parameter is chosen here so as to obtain a finite value of $X(S;m_0,n_0)$ for any value of $S$. This occurs when $M>S^{max}-S^{min}=Z$. Apparently, $m_0,n_0\geq 1$ is the natural lower limit of $m_0,n_0$, because only then $X(S;m_0,n_0)$ disappears for $S=S^{min},S^{max}$, like this should be. We still assume more strongly that $m_0,n_0\geq 2$, which has already been mentioned above. We emphasize that $X$ does not scale with $N$ as opposed to partial measures of complexity.

Note that for $M \gg Z$, both measures of complexity have approximate values $X(S; m0, n0)\approx CX(S; m0, n0)$. Important differences between these two measures only appear for $Z/M$ close to 1, because only then the denominator in Eq. (\ref{rown:CXmnwN}) plays an important role. Of course, $M$ is a free parameter, and possibly its specific value could be obtained from some additional (e.g., external) conditions.

\begin{figure}[H]
\centering
\includegraphics[width=10cm]{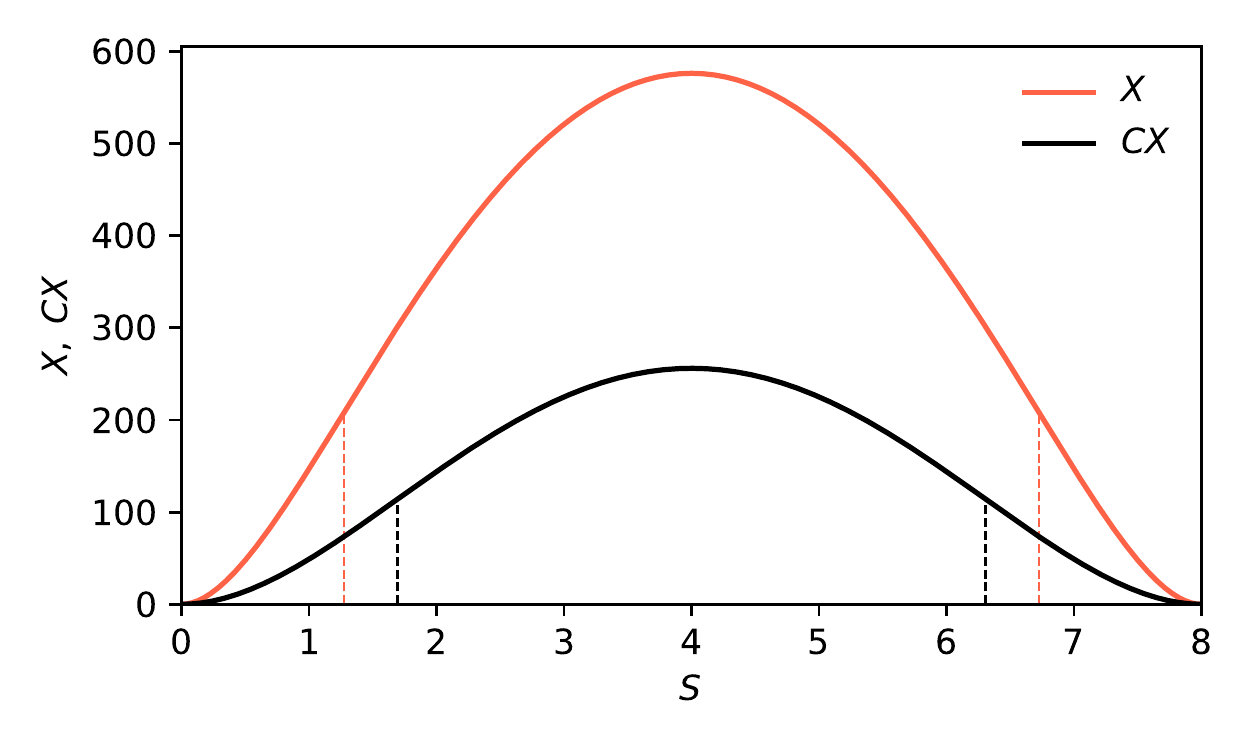}
\caption{Comparison of the partial measure of complexity $CX(S;m=2,n=2)$ and full measure of  complexity $X(S;m_0=2,n_0=2)$ for the symmetric case of $m=n=m_0=n_0$. In addition, we assumed that $S^{min}=0, S^{max}=8$ and $M=10$ for both measures. For both cases Eqs. (\ref{rown:SCX2}), (\ref{rown:CX2}) and (\ref{rown:CXmnwN}) are valid. Vertical dashed lines indicate inflection points: black for the $CX$ curve, orange for the $X$ curve, while $S_{CX}^{max}=S_{X}^{max}=4$.}
\label{figure:FIG_4_extra}
\end{figure}

In Fig. \ref{figure:FIG_4_extra} we compare the behavior of the partial (black curve) and full measure (orange curve) of complexity, where we used the entropy instead of the specific entropy. Whether $CX$ lies below or above $X$ depends both on $M$ parameter (determining the weight at which individual measures of partial complexity enter the full measure of complexity), as well as on the $Z/M$ ratio. 

We continue to determine the entropic susceptibility of the full measure of complexity,
\begin{eqnarray}
\chi _X&=&\frac{dX(S;m_0,n_0)}{dS}=(m_0+n_0)(S_{CX}^{max}-S)\chi_x (S;m_0-1,n_0-1)\nonumber \\
&+&\frac{2}{M^2}X(S;m_0,n_0)\frac{S-S^{arit}}{(1-\frac{S^{max}-S}{M})(1-\frac{S-S^{min}}{M})}.
\label{rown:dXS}
\end{eqnarray}
\begin{figure}[H]
\centering
\includegraphics[width=10cm]{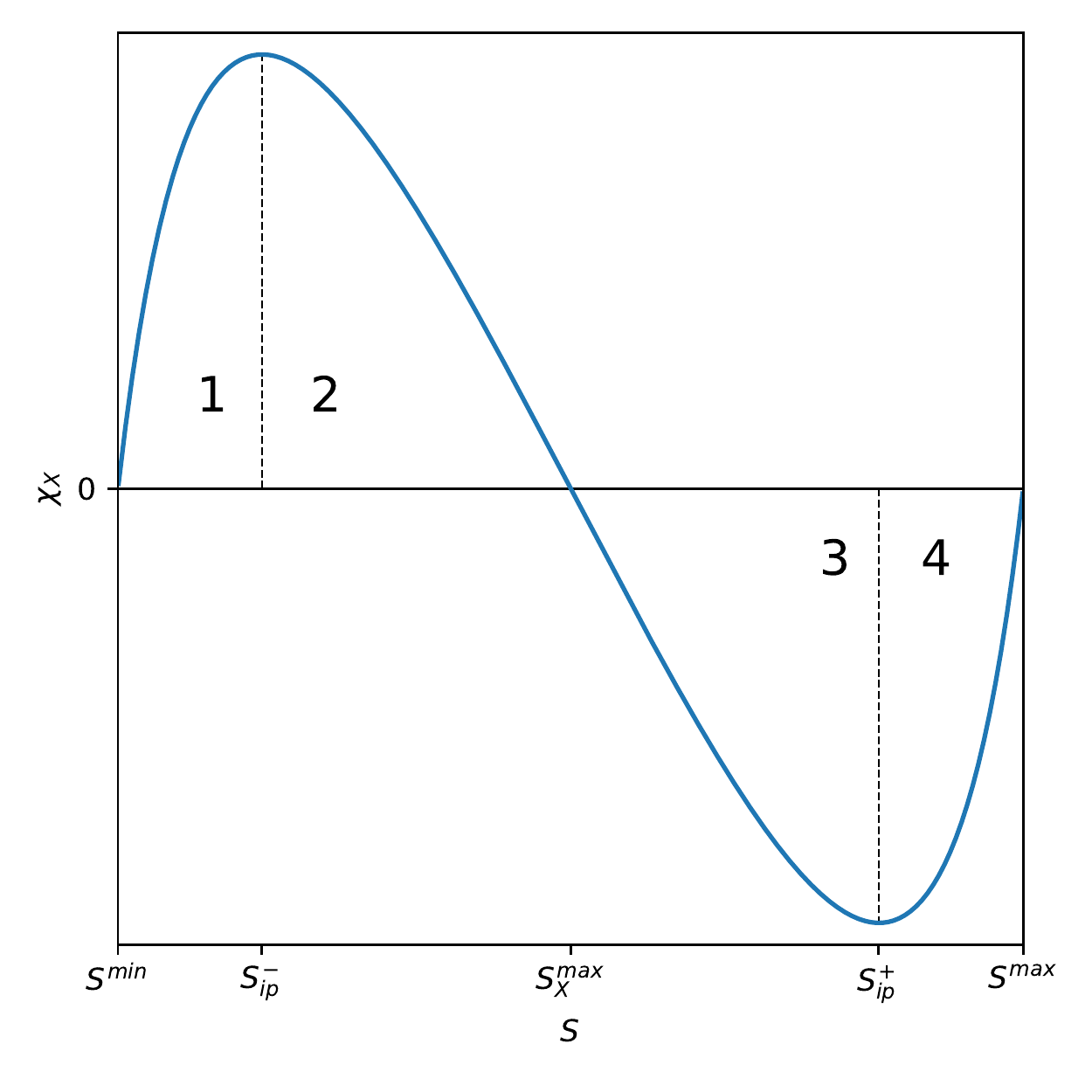}
\caption{Schematic plot of the entropic (non-equilibrium) susceptibility $\chi_{X}(S;m,n)$ of the full measure of complexity vs. $S$ at arbitrary fixed order $(m_0=2,n_0=2)$. As expected from Fig. \ref{figure:FIG_4_extra}, turning points of $CX$ (cf. Fig. \ref{figure:Suscepti}) lie within the $S$ interval bounded by inflection points of $X$.}
\label{figure:FIG_5}
\end{figure} 

Thus, the evolution of $X$ is governed by an equation analogous to Eq. (\ref{rown:CXSt}) except that $\chi _{CX} $ present in this equation should be replaced by $\chi _X $ given by Eq. (\ref{rown:dXS}).

\subsection{Distribution non-equilibrium entropies}\label{section:entropy}

Distribution entropy is understood as the entropy based on coarse-grained probability distributions -- this type of entropy is most often used \cite{Kamp,BeckSchl,LLZh}. A very characteristic example is the entropy class built on time-dependent probability distributions, $\{p_j(t)\}$, satisfying master (Markovian), or M-equation, in presence of detailed balance conditions. We give here two very characteristic (nonequivalent) examples of this type of entropies\footnote{More specifically: one should talk about specific entropy. However, we continue to leave the word 'specific,' as this does not lead to confusion in this work. In addition, entropies given by Equations (22) and (23) belong to the category of relative entropies. We can call them generalized Kullback-Leibler entropies \cite{PBEH}.} 
\begin{eqnarray}
S(t)=S_0\left[1-\sum_{j}p_{j}^{eq}f\left(\frac{p_j(t)}{p_{j}^{eq}}\right)\right] 
\label{rown:entrop1}
\end{eqnarray}
and
\begin{eqnarray}
S(t)=-S_0\ln{\sum_{j}p_{j}^{eq}f\left(\frac{p_j(t)}{p_{j}^{eq}}\right)},
\label{rown:entrop2}
\end{eqnarray}
where $p_j(t)$ is a probability of finding a system in state $j$ at time $t$, while $p_j^{eq}$ is a corresponding equilibrium probability. We are considering only discrete states here. The function $S_0f(x)\geq 0$, where domain $0\leq x\leq \infty $, is a non-negative convex function obeying $S_0\frac{d^2f}{dx^2}\geq 0$. It can be shown \cite{Kamp} that entropies defined in this way meet the law of entropy increase, i.e. its derivative
\begin{eqnarray}
\frac{dS(t)}{dt}\geq 0; 
\label{rown:st0}
\end{eqnarray}
therefore $S(t)\rightarrow S^{max}$ from below when $p_j(t)\rightarrow p_j^{eq}$. Eq. (\ref{rown:st0}) is the key property of entropy. Let us add that for $p=p^{eq}$ entropy defined by equations (\ref{rown:entrop1}) and (\ref{rown:entrop2}) disappear. In other words, these entropies are negative and grow to zero as the system tends to equilibrium. 

It is worth paying attention to the possibility of defining generalized information gain, whereby this information gain is calculated here relative to the equilibrium distribution. We can write,
\begin{eqnarray}
\Delta I(p(t),p^{eq})=-S(t),
\label{rown:Kull}
\end{eqnarray}
where $p(t)=\{p_j(t)\}$ and $p^{eq}(t)=\{p_j^{eq}(t)\}$. Furthermore, entropy $S(t)$ is closely related to partition function. Therefore, in this approach, the entropy is a base function.
 
 Most often the function $f(x)$ is selected in the form,
 \begin{eqnarray}
 f(x)=x^{\alpha },
 \label{rown:fx}
 \end{eqnarray}
 coupled with a constant $S_0=\frac{1}{\alpha -1}$. With these choices the entropy given by Eq. (\ref{rown:entrop1}) is called Tsallis entropy and the entropy given by Eq. (\ref{rown:entrop2}) R\'enyi entropy. Usually, the entropic index $\alpha $ is denoted by $q$ in the case of Tsallis entropy. 
 
 Both Tsallis and R\'enyi entropies converge to Boltzmann-Gibbs-Shannon (BGS) entropy\footnote{More precisely, entropies given by formulas (\ref{rown:entrop1}) and (\ref{rown:entrop2}) converge to Kullback-Leibler relative entropy. In our case, this entropy, we calculate relative to the state of thermodynamic equilibrium of the system.} when the entropy index tends to 1. We remind that the BGS entropy has two basic features: 
 \begin{itemize}
     \item[(i)] obeys the Boltzmann $H$-theorem (for dilute binary interacting gases),
     \item[(ii)] it is an additive quantity that becomes extensive for a gas or a solid.
 \end{itemize}
The Tsallis and R\'enyi entropies are particularly useful that can be used in both extensive/linear, non-extensive/nonlinear, as well as equilibrium/non-equilibrium physics and also beyond the physics domain. 

\section{Finger print of complexity in simplicity}\label{section:idgas}

Let's consider the perfect gas at a fixed temperature which is initially closed in the left half of an isolated container. Then the partition next is removed, and the gas undergoes a spontaneous expansion. We are dealing here (practically speaking) with an irreversible process even for a small number of particles (at least the order of $10^2$).

Let's recall the definition of 'perfect gas.' It is a gas of particles that cannot `see' each other, i.e., there are no interactions between them. Thus, from a physical point of view, it is a dilute gas at high temperature. We further assume that all particles have the same kinetic energy. The legitimate question is whether such a gas will expand after the partition is removes? We notice that the thermodynamic force is at work here, being roughly proportional to the difference in the number of particles in the right and left parts of the container. This force causes the expansion process. Thus, we are dealing with the simplest paradigmatic irreversible process \cite{KTH}. The particles remain stuck in the final state and will not leave it (with accuracy to slight fluctuations in the number of particles in the right half of the container). Such a final state of the whole system is referred to as the equilibrium state. The simple coarse-grain description of the system allows us to introduce here the concept of configuration entropy.

Note that the macroscopic state of the system (generally, the non-equilibrium one) can be described by the instantaneous number of particles in the left ($N_L$) and right ($N_R$) parts of the container, with $N=N_L+N_R$, where $N$ is the total fixed number of particles. It allows one to define the weight of the macroscopic state $\Gamma (N_L)$, also called thermodynamic probability. This is the number of ways to arrange the $N_L$ particles in the left part of the container and $N_R=N-N_L$ in the right. Hence,
\begin{eqnarray}
\Gamma (N_L)=\frac{N!}{N_L!(N-N_L)!}.
\label{rown:Gama}
\end{eqnarray}
We do not distinguish here permutations of particles inside each part of the container separately. We take into account only permutations of particles located in different halves of the container. This is because our resolution is too small here to observe the location of particles inside each container separately. Such a coarse-graining creates an information barrier: more information can mask the complexity of the system. We will not be able to see the complexity because we will not be able to construct entropy. This creates a paradoxical situation: the surplus of information makes the task difficult and does not facilitate obtaining the insight into the system. Here we have an analogy with chaotic dynamics, where chaos is visible only in the Poincare surface cross-section of the phase space and not in the entire phase space.

The configuration entropy at a given time $t$ we define as follows,
\begin{eqnarray}
S(N_L(t))=\ln \Gamma (N_L(t)),
\label{rown:St}
\end{eqnarray}
where $\Gamma (N_L)$ is given by Eq. (\ref{rown:Gama}). The above expression can be used both for the equilibrium and non-equilibrium states.

It can be demonstrate using the Stirling formula that for large $N$, entropy is reduced to the BGS form,
\begin{eqnarray}
\ln \Gamma (N_L)=-N\left[p_L(t)\ln p_L(t) + p_R(t)\ln p_R(t) \right]=Ns(t),
\label{rown:SNt}
\end{eqnarray}
where $p_J(t)\stackrel{\rm def.}{=}\frac{N_J(t)}{N},~J=L,R$, and $s(t)$ is a specific entropy. The law of entropy increase (\ref{rown:st0}) is also fulfilled here, as expected.

We now obtain the equation for determining $N_L^{CX^ {max}}$, i.e. the number of particles in the left part of the container that maximizes the partial complexity measure $CX$. To this end, we substitute $N_L=N_L^{CX^{max}}$ into the right side of Eq. (\ref{rown:St}) and in Eq. (\ref{rown:SCX2}) we put $S^{max}$ equal to the right-hand side of Eq. (\ref{rown:St}) for $N_L=N/2=N^{eq}$. Hence, we obtain a constitutive equation,
\begin{eqnarray}
S(N_L=N_L^{{CX}^{max}})=S_{CX}^{max},
\label{rown:constit}
\end{eqnarray}
where $N_L^{{CX}^{max}}$ is our sought quantity. We assumed to the right of the above equation (as it is commonly used), that $S^{min}$ = 0. Thus, only the relation $S^{max}=S(N^{eq})$ is taken into account. Equation (\ref{rown:constit}) is an example of Eq. (\ref{rown:constitS}), where $N_L^{CX^{max}}$ plays the role of $Y^{CX^{max}}$. This equation has the following explicit form,
\begin{eqnarray}
\left[\Pi_{j=1}^{N-N_L^{CX^{max}}}\left(1+\frac{N_L^{CX^{max}}}{j}\right)\right]^2=\Pi_{j=1}^{N/2}\left(1+\frac{N}{2j}\right), ~\mbox{for $n=m=2$}.
\label{rown:Pi}
\end{eqnarray}
Just deriving Eq. (\ref{rown:Pi}) was the primary purpose of this example.
This is a transcendental equation which exact analytical solution is unknown. When deriving Eq. (\ref{rown:Pi}) we used the initial condition for the entropy that is, $S(t=0)=S^{min}=\ln \Gamma (N_L=N)=0$, which follows from Eqs. (\ref{rown:Gama}) and (\ref{rown:St}). Even for such a simple toy model, determining the partial measure of complexity is a non-trivial task, also because $N_L$ is different from $N/2$ (as we show below). 

\begin{figure}[H]
\centering
\includegraphics[width=10cm]{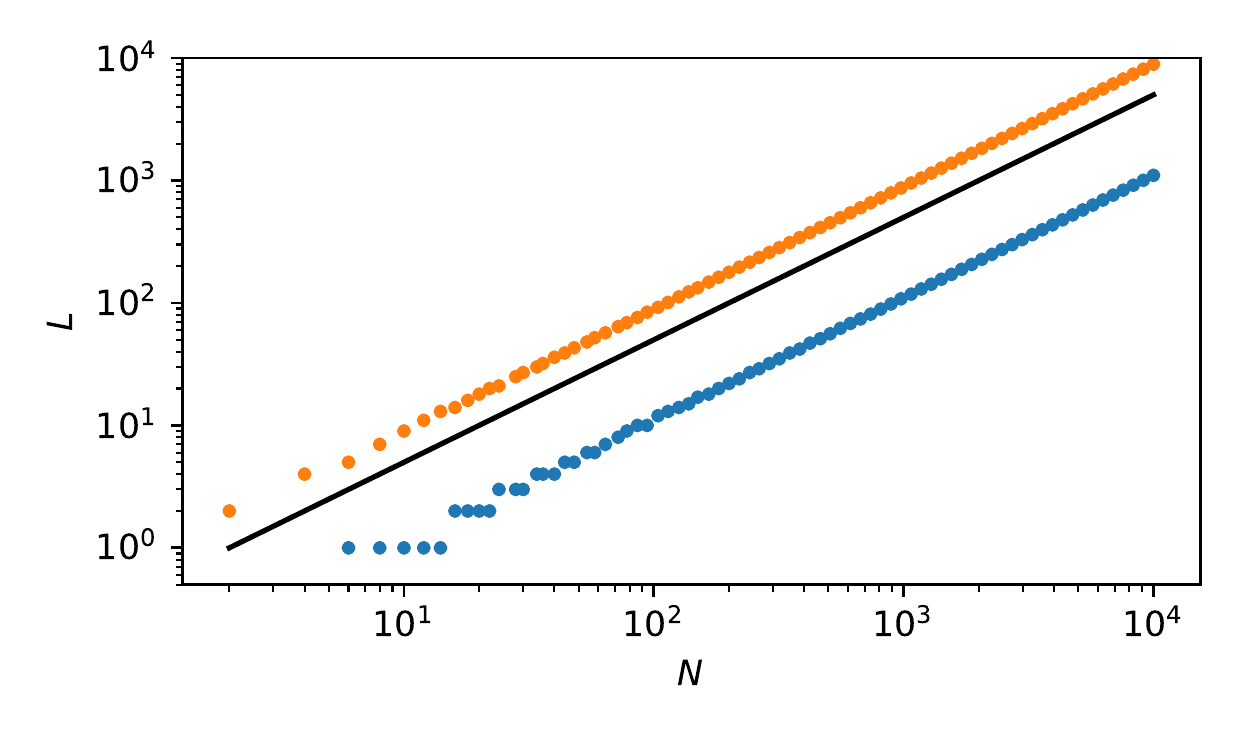}
\caption{Dependence of $L(=N_L^{CX^{max}})$ vs. $N$. There are two solutions of Eq. (\ref{rown:Pi}): one marked with blue circles and the other with orange ones. Above $N\approx 10^2$, both dependencies are linear, which is particularly clearly confirmed in Fig. ( \ref{figure:LNdir}). That is, in a log-log scale, their slopes equal 1.  However, in linear scale, the directional coefficients of these straight lines equal $0.11$ and $0.89$, respectively. It is clearly shown in Fig. \ref{figure:LNdir}. Only the solution with orange circles is realistic, because the chance that $89\% $ of particles will pass in the finite time to the second part of the container (as indicated by the solution marked with blue circles) is negligibly small The black solid tangent straight line indicates a reference case $N_L^{CX^{max}}=N/2$.}
\label{figure:LN}
\end{figure}

The numerical solutions of Eq. (\ref{rown:Pi}), i.e. the relationship of $N_L^{CX^{max}}$ to $N$, are shown in Fig. \ref{figure:LN} (for simplicity, $L$ defining the vertical axis on the plot means $N_L^{CX^{max}}$). Both solutions (small circles above and below the solid straight line) show that $N_L^{CX^{max}}$ is significantly different from $N/2$. Thus, the most complex state is significantly different from the equilibrium state.
\begin{figure}[H]
\centering
\includegraphics[width=10cm]{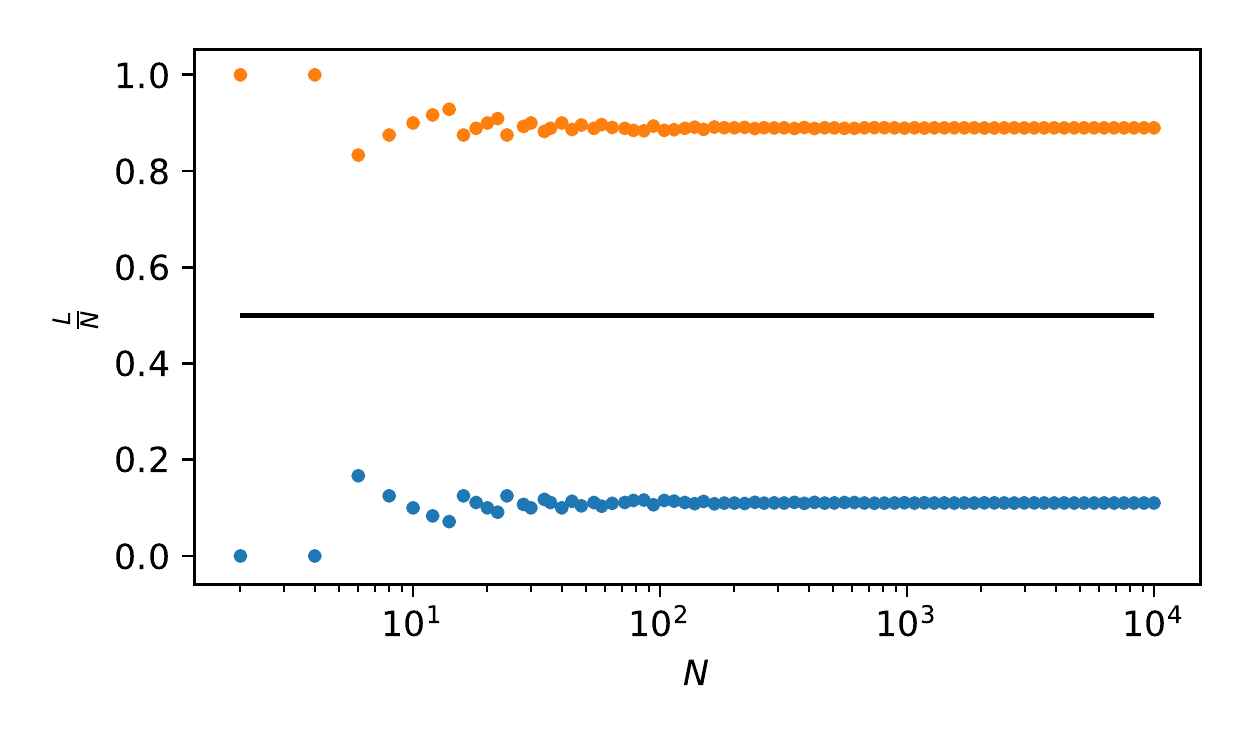}
\caption{Directional coefficient of linear dependencies $L$ vs.$N$ as a function o $N$. For $N$ greater than $10^2$, no $N$-dependence of this coefficient is observed. Both solutions (having $L/N=0.11$ and $L/N=0.89$) are mutually symmetric about the straight horizontal line $L/N=1/2$ but only the solution $L/N=0.89$ we consider as realistic. The black horizontal straight solid line indicates a reference case $N_L^{CX^{max}}=N/2$.}
\label{figure:LNdir}
\end{figure}

Having the $N_L^{{CX}^{max}}$ dependence on $N$, we obtain the dependence of complexity $CX^{Xmax}$ on $N$ order $(m=2,n=2)$. We can write,
\begin{eqnarray}
CX^{max}=\left(\frac{S(N/2)}{2}\right)^4=\left[\frac{1}{2N}\ln\left( \Pi_{j=1}^{N/2}\left(1+\frac{N}{2j}\right)\right)\right]^4,
\label{rown:CXmaxgas}
\end{eqnarray}
as in our case $S^{max}=S(N/2)$ equals the logarithm of the right-hand side of Eq. (\ref{rown:Pi}) divided by $N$. Notably, Eq. (\ref{rown:CXmaxgas}) is based on Eq. (\ref{rown:CX2}). 

\begin{figure}[H]
\centering
\includegraphics[width=10cm]{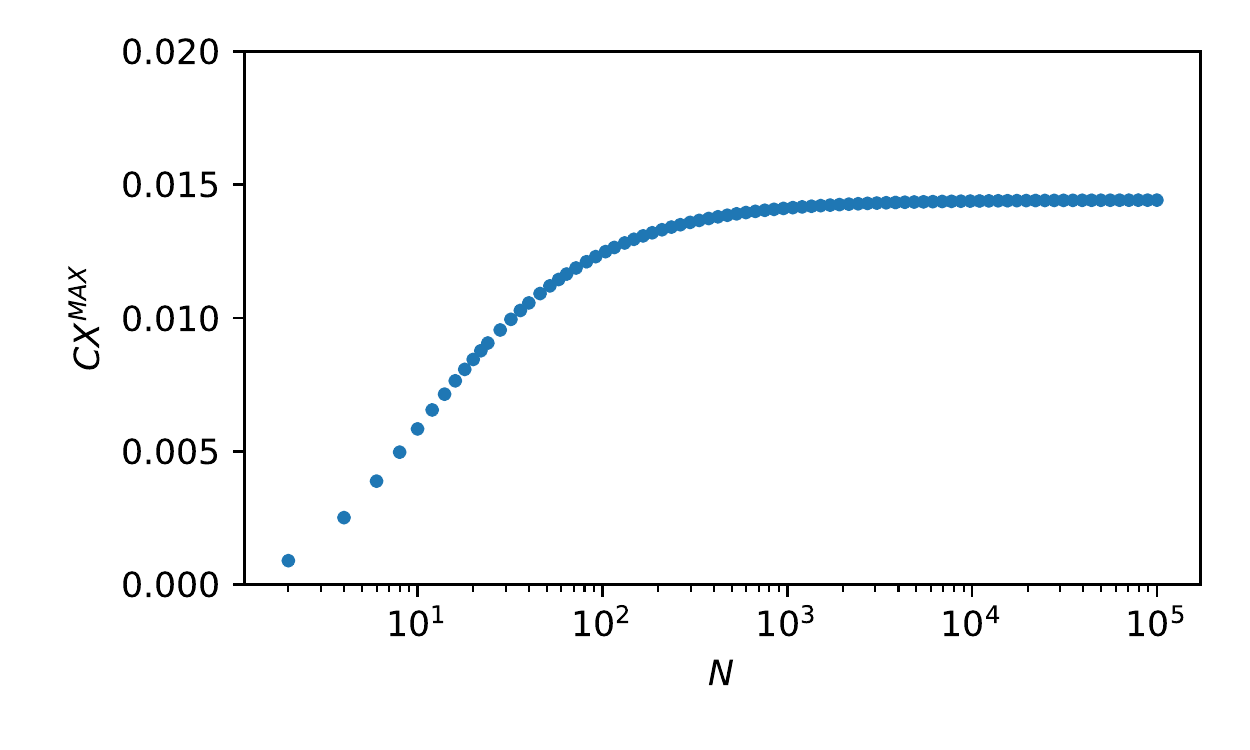}
\caption{Dependence of $CX^{max}$ on $N$. As one can see, $CX^{max}$ is a non-extensive function -- it reaches the plateau for $N\gg 1$. For $ N\approx 10^4 $ the plateau is achieved with a good approximation. This is an important issue for researching complexity. Namely, systems can attain complexity already on a mesoscopic scale.}
\label{figure:CXmaxN}
\end{figure}

In Fig. \ref{figure:CXmaxN} we present the dependence of $CX^{max}$ on $N$. $CX^{max}$ is a non-extensive function -- it reaches the plateau for $N\gg 1$. For $ N\approx 10^4 $ the plateau is achieved with a good approximation. It can be said that $CX^{max}$ behaves like specific complexity measure. This is important for researching complexity. Namely, systems can attain complexity already on a mesoscopic scale. Although the absolute value of the complexity measure is relatively small, it is evident and possesses a structure related to the current inflection point there (near $N=10$). 

This example shows that even such a simple arrangement of non-interacting objects may have non-disappearing non-equilibrium complexity. A necessary (but not sufficient) condition is the possibility of constructing entropy and the presence of a time arrow.

\section{Concluding remarks}\label{section:conclus}

In many recent publications~\cite{Pincus,PBEH,RichMo,LLZh} it is argued that entropy can be a direct measure of complexity. Namely, the smaller value of entropy indicates more regularity or lower system complexity, while its larger value indicates more disorder, randomness and higher system complexity. However, according to Gell-Mann, a more disorder means less and not more system complexity. These two viewpoints are contradictory -- this is a serious problem, which we addressed.

Our motivation in solving the above problem was based on the Gell-Mann's view of complexity. This is because fail to agree that the loss of information by the system as it approaches equilibrium, increases its complexity; notably, $\Delta I(p^{eq},p^{eq})$ takes its minimum value then and complexity must decrease. 
In addition, the differences in definition (\ref{rown:CXS}) eliminate the useless dependence of complexity on the additive constant that may appear in the definition of entropy. It can be said that the system state with the highest complexity is the most state most distant from all the states of the system of lesser or no complexity.

Thus, in the sense of Gell-Mann, the measure of complexity should supply a complementary information to the entropy or its monotonic mapping. 

Therefore, in this work, we presented a methodology which allows building a universal measure of complexity as a function of system state based on the non-linearly transformed entropy. This is a non-extensive measure. This measure should meet a number of conditions/axioms that we indicated at work. A parsimonious example, of the simplest system with a small and a large number of degrees of freedom,is presented to support our methodology. As a result of this approach, we have shown that (generally speaking) the most complex are optimally mixed states consisting of pure states, i.e., of the most regular and most disordered, which the space of states of a given system allows.

We should pay attention to an essential issue regarding the definition of the phenomenological partial measure of complexity given by definition (\ref{rown:CXS}). This definition is open in the sense that if the description of complexity requires, for example, one additional quantity, then the definition (\ref{rown:CXS}) takes on an extended form,
\begin{eqnarray}
CX(S,E;m_1,n_1,m_2,n_2)\stackrel{\rm def.}{=}(S^{max} - S)^{m_1}(S-S^{min})^{n_1}(E^{max}-E)^{m_2}(E-E^{min})^{n_2} \geq 0, 
\label{rown:def2D}
\end{eqnarray}
whereby $E^{min}\leq E\leq E^{max}$ this new quantity was marked. This definition has still an open character. Specifically, this definition also allows (if the situation requires it) to replace one quantity with another, e.g., entropy with free energy or considering some derivatives (e.g., of the type $\frac{\partial S}{\partial E}$). Openness and substitutability should be the key features of the measure of complexity. What's more, exponents $ m_j,~n_j,~j = 1,2,$ determine the order of complexity, i.e. its level or scale. We emphasize that the introduced measure of complexity can describe isolated and closed systems (although in contact with the reservoir), as well as open systems that can change their elements.

From Eqs. (\ref{rown:def2D}) and (\ref{rown:umsX}) we get the phenomenological universal full measure of complexity in the form which extends Eq. (\ref{rown:CXmnwN}),
\begin{eqnarray}
X(S;m_1^0,n_1^0,m_2^0,n_2^0)&=&\left(1-\frac{1}{M_1}\right)^2\frac{(S^{max}-S)^{m_1^0}}{1-\frac{S^{max}-S}{M_1}}\frac{(S^{min}-S)^{n_1^0}}{1-\frac{S^{min}-S}{M_1}} \nonumber \\
&\times &\left(1-\frac{1}{M_2}\right)^2\frac{(E^{max}-E)^{m_2^0}}{1-\frac{E^{max}-E}{M_2}}\frac{(E^{min}-E)^{n_2^0}}{1-\frac{E^{min}-E}{M_2}}\ge 0.
\label{rown:extend_def21}
\end{eqnarray}

Definitions of measures of complexity (\ref{rown:CXS}) and (\ref{rown:CXmnwN}) and their possible extensions are universal and useful. It is due to entropy associated not only with thermodynamics (Carnot, Clausius, Kelvin) and statistical physics (Boltzmann, Gibbs, Planck, R\'enyi, Tsallis) but also with the information approach (Shannon, Kolmogorov, Lapunov, Takens, Grassberger, Hantschel, Procaccia) and with the approach from the side of cellular automata (von Neumann, Ulam, Turing, Conway, Wolfram, et al.), i.e., with any representation of the real world using a binary string. Today, we have already several very effective methods for counting entropy of such strings as well as other macroscopic characteristics sensitive to the organization and self-organization of systems, as well as to their synchronization (synergy, coherence), competition, reproduction, adaptation -- all of them sometimes having local and sometimes global characters. 

Our definitions of complexity also goes out to meet the research of the complexity of the biologically active matter. In this, especially research on the consciousness of the human brain can get a fresh impulse. The point is that most researchers believe that the main feature of conscious action is a maximum complexity, i.e., critical complexity \cite{Sornett} -- in our approach it would be $CX^{max}$. It separates the phase with the predominance of ordered states from the one containing the predominance of disordered states, as it should be. However, to achieve this strictly, one would have to go from positive integers $m$ and $n$ to positive continuous exponents, $\alpha $ and $\beta $, respectively. The criticality would be recoverable for $\alpha $ or $\beta $ smaller than 1. Then it would be possible to show the singular behavior of the entropic susceptibility. As such, our approach might serve to study the evolution/dynamics of consciousness.

Therefore, we hope that our approach will enable (i) the universal classification of complexity, (ii) analysis of a system critical behavior and its applications, and (iii) study the dynamic complexity. All these constitute create the background of science of complexity.\\
\newline
\textbf{Author contributions}: RK and ZRS conceptualised the work,; RK wrote the draft and conducted the formal analysis and prepared the draft of figures; JK provided numerical calculations and provided the final figures; ZRS finalised the manuscript. All authors read and approved the final manuscript.\\
\newline
\textbf{Funding}: One of the authors of the work (ZRS) benefited from the financial support of the ZIP Program. This program of integrated development activities of the University of Warsaw is implemented under the operational program Knowledge Education Development, priority axis III. Higher education for economy and development, action: 3.5 Comprehensive university programs, from April 2, 2018 to March 31, 2022, based on the contract signed between the University of Warsaw and the National Center for Research and Development. The program is co-financed by the European Union from the European Social Fund; \url{http://zip.uw.edu.pl/node/192}. Apart from that, there was no other financial support. \\
\newline
\textbf{Conflicts of interest}: The authors declare no conflict of interest.

\bibliographystyle{unsrt}  


\end{document}